\documentclass[preprint,12pt]{elsarticle}

\usepackage{amsmath,amssymb,amsthm}
\usepackage{booktabs}
\usepackage{array}
\usepackage{longtable}
\usepackage{graphicx}
\usepackage{multirow}
\usepackage{hyperref}
\usepackage{url}
\usepackage{listings}
\usepackage{xcolor}
\usepackage[margin=1in]{geometry}
\usepackage{tabularx}
\usepackage{siunitx}

\journal{Preprint prepared in Elsevier \texttt{elsarticle} style}

\begin{document}

\begin{frontmatter}

\title{Beyond TVL: An Explainable Risk Scoring Framework for Tokenized Real-World Assets}

\author{Rischan Mafrur, Khadijah}
\address{Western Sydney University, \href{https://cryptocoinhalal.com/}{CryptoCoinHalal} \\ R.Mafrur@westernsydney.edu.au , Khadijah@cryptocoinhalal.com}

\begin{abstract}
Tokenized real-world assets (RWAs) are often evaluated through headline indicators such as total value locked (TVL) or on-chain asset value. However, a large asset base does not necessarily imply low risk, since tokenized assets may remain illiquid, weakly traded, or highly concentrated among a small number of holders. Using public data from RWA.xyz, this paper develops an empirical and explainable risk scoring framework for tokenized RWA markets. The framework evaluates three dimensions of risk: liquidity risk $L$, concentration risk $C$, and market-quality risk $M$. These risk dimensions are constructed from observable indicators, including turnover, holder distribution, active-address activity, transfer frequency, and network concentration measured through Herfindahl indices. The analysis shows that several RWA tokens with substantial on-chain value exhibit high empirical risk because they combine limited transfer activity, low turnover, and concentrated ownership structures. In contrast, assets with broader participation and stronger on-chain activity display lower liquidity and concentration risk, even when their headline asset values are smaller. The findings demonstrate that TVL alone can obscure important risks in tokenized asset markets. By providing a transparent and data-driven risk scoring approach, this paper contributes to the empirical assessment of RWA liquidity and offers a practical basis for comparing tokenized assets beyond headline valuation metrics.
\end{abstract}


\end{frontmatter}

\section{Introduction}

Tokenized real-world assets are moving from a conceptual promise to a measurable market segment. Policymakers and market participants increasingly view tokenisation as a potentially important infrastructure layer for money, securities, funds, and collateral management \citep{BIS2023Continuum,CPMI2024Tokenisation,IMF2026TokenizedFinance}. At the same time, official reports stress that tokenisation does not remove the economic, legal, or operational frictions embedded in the underlying claim structure \citep{IOSCO2025Tokenization,IMF2025Inefficiencies}. In other words, putting an asset on-chain does not automatically make it liquid, transparent, or robust.

This point matters acutely for tokenized real-world assets. In market practice, the quality of tokenized RWAs is often summarised with scale metrics such as TVL, market value, assets under management, or circulating supply. Those indicators are useful for measuring adoption, but they are not equivalent to exitability. Market microstructure has long shown that liquidity is distinct from size, and that depth, breadth, fragmentation, and price impact matter for execution quality \citep{Kyle1985,Amihud2002,OHaraYe2011}. Information economics further suggests that where asset quality is difficult to verify, disclosure quality and credible certification become central to market functioning \citep{Akerlof1970}. For tokenized RWAs, those concerns are amplified by the coexistence of on-chain transferability and off-chain reserve, custody, legal, and redemption processes.

The empirical challenge is therefore not whether tokenized RWAs are ``large'', but whether they are structurally investable. A tokenized treasury fund with billions of dollars in value may still be unsuitable for many investors if it has only a few dozen holders, a narrow dealing window, a high minimum redemption amount, or uncertain legal segregation. Conversely, a somewhat smaller product may be materially more investable if it combines wider holder breadth, frequent reporting, stronger legal form, lower entry barriers, and more consistent activity.

To address this, we propose an \emph{explainable risk scoring framework} focusing on liquidity, concentration, and market-quality dimensions.  Each dimension is constructed from publicly measurable RWA.xyz data (holders, transfers, active addresses, network distribution).  We compute (a) {\it liquidity risk} based on turnover and transfer activity; (b) {\it concentration risk} based on holder breadth and wealth-per-holder; and (c) {\it market-quality risk} based on cross-chain participation and trading concentration.  Each component is normalized and averaged for a composite risk score.  Our pilot sample of 10 tokenized RWAs (U.S. Treasury funds, credit funds, and gold tokens) illustrates how assets with similar TVL can have very different risk profiles under the L--C--M scores.

The key contribution is one of the first empirical, data-driven frameworks to rank tokenized RWAs by observable liquidity and activity metrics.  We emphasize that TVL and static size metrics are insufficient on their own: the \emph{observable} ease of exiting (turnover and active participation) and ownership dispersion are crucial for assessing investability.  By using only public RWA.xyz data, the resulting $L,C,M$ scores are transparent, reproducible, and explainable.  We show examples where large funds with many holders have low liquidity risk, while smaller funds with concentrated ownership can score worse.  This work targets practitioners and researchers who need to go beyond headline TVL and understand the actual secondary-market dynamics of tokenized assets.  

\section{Literature review}

\subsection{Tokenisation and financial market structure}

Recent institutional work treats tokenisation as a potentially important redesign of financial market plumbing rather than a mere repackaging of cryptoassets \citep{BIS2023Continuum,CPMI2024Tokenisation,IMF2026TokenizedFinance}. The IMF argues that tokenisation may reduce frictions across parts of the asset life cycle, but that the benefits depend on the persistence of market inefficiencies, interoperability limits, and legal design \citep{IMF2025Inefficiencies}. IOSCO likewise emphasises that investor protection, transfer validity, settlement finality, and legal enforceability remain central risks in tokenized financial assets \citep{IOSCO2025Tokenization}.

These observations are especially relevant for RWAs because tokenisation creates a layered claim. The token is on a public or semi-public ledger, but the asset, the reserve, the bankruptcy waterfall, and redemption processes remain partly or wholly off-chain. As BIS has argued, tokenisation sits on a ``continuum'' in which technical feasibility is only one dimension; legal recognition and operational trust remain binding constraints \citep{BIS2023Continuum}.

\subsection{Liquidity, exitability, and market quality}

Classical market microstructure distinguishes capitalisation from liquidity. Kyle's model links liquidity to price impact and informed trading \citep{Kyle1985}; Amihud formalises an illiquidity premium using price impact relative to trading activity \citep{Amihud2002}; O'Hara and Ye show that fragmentation can affect market quality depending on how order flow and liquidity are distributed \citep{OHaraYe2011}. The implication for tokenized RWAs is direct: a large tokenized fund may still be difficult to exit if activity is episodic, fragmented, or dominated by internal rebalancing rather than organic two-sided trading \citep{mafrur2025tokenize}.

BIS work on tokenized money market funds reinforces this point. Tokenized MMFs can mirror and in some cases amplify traditional liquidity mismatch and run-dynamics concerns, particularly when they are used as collateral or settlement assets in layered market structures \citep{BIS2025TMMF}. Related BIS work on tokenized real estate similarly cautions against assuming that tokenisation alone creates deep secondary markets \citep{BIS2025RealEstate}. In short, tokenisation does not automatically create liquidity.

\section{Research gap and contributions}

Existing RWA analyses are mostly descriptive or case-based.  Academic studies focus on specific domains (e.g.\ tokenized real estate) and find generally low trading \citep{swinkels2025empirical}.  However, there has been little prior work on a general, cross-asset risk rating methodology for tokenized RWAs.  In particular, we are aware of no published framework that (a) uses public data to score RWA liquidity and concentration, and (b) produces an explainable index combining multiple dimensions.  This paper fills that gap.

Our contributions are: (1) We develop an explainable, empirical $L$-$C$-$M$ scoring framework for tokenized RWAs.  The dimensions are chosen for data availability and theoretical relevance.  (2) We illustrate the framework on 10 diverse RWA tokens using real data from RWA.xyz. (3) We demonstrate that widely reported figures (TVL, market cap) can be misleading: e.g.\ a token with \$2B in assets may have turnover and trading risk akin to a token with only \$10M market cap if active trading is negligible. (4) We provide sensitivity analysis, showing how the composite score changes if an investor weights liquidity or concentration more heavily.

The goal is not to replace expert risk analysis, but to provide a transparent quantitative screening tool.  All data used are public and verifiable.  We explicitly discuss limitations (e.g.\ unobserved off-chain trading, legal risk) and treat the framework as a pilot for further research.

\section{Data and methodology}

\subsection{Sample selection}

We use publicly accessible data from RWA.xyz, an analytics platform for tokenized assets.  RWA.xyz aggregates on-chain and issuer-provided data for every listed RWA, including asset value, token supply, holders, and transfer activity.  We select 10 representative tokens spanning categories: five U.S. Treasury funds (\texttt{BUIDL}, \texttt{BENJI}, \texttt{OUSG}, \texttt{USTB}, \texttt{USDY}), one major stablecoin (\texttt{USDC}), two private-credit funds (\texttt{HLSCOPE}, \texttt{STAC}), and two gold tokens (\texttt{PAXG}, \texttt{XAUT}). These were chosen for high on-chain value and category diversity which explained in detail in Table~\ref{tab:sample}.
\begin{table}[htbp]
\centering
\caption{Pilot sample of tokenized RWAs}
\label{tab:sample}
\scriptsize
\setlength{\tabcolsep}{3pt}
\renewcommand{\arraystretch}{1.15}
\begin{tabularx}{\textwidth}{lXXXX}
\toprule
Ticker & Category & Token form & Public access profile & Illustrative structural feature \\
\midrule
BUIDL & Tokenized treasury/MMF & Fund share token & U.S. qualified purchasers & Daily dealing, but very high minimum primary-market size \\
BENJI & Tokenized treasury/MMF & Mutual fund share token & U.S. retail and institutional investors & Registered fund structure with low minimum investment \\
OUSG & Tokenized treasury & Fund token & U.S. qualified purchasers & Daily and instant redemption architecture, qualified access only \\
USTB & Tokenized treasury & Fund token & Accredited and qualified purchasers & Banking-day dealing and upgradable contract design \\
USYC & Tokenized treasury/MMF & Fund share token & Non-U.S. investors & Large asset value with narrow holder base \\
USDY & Yield-bearing treasury-linked note & Tokenized note & Qualifying non-U.S. investors & Strong activity but meaningful legal and eligibility restrictions \\
HLSCOPE & Private credit & Tokenized feeder fund share & Accredited investors & Monthly subscription and redemption cycle \\
STAC & Structured credit & Tokenized fund share & Reg D / Reg S eligible investors & Very concentrated holder base and explicit redemption fee \\
PAXG & Gold & Commodity-backed token & Retail and institutional access & Strong breadth and activity; redemption and custody remain central \\
XAUT & Gold & Commodity-backed token & Non-U.S. investors & High redemption minimum despite strong transfer activity \\
\bottomrule
\end{tabularx}
\end{table}

For each token, we extract all visible variables on its RWA.xyz asset page as of May 2026.  Specifically, we collect: ticker, asset name, category, total asset value (on-chain market cap or NAV), net asset value per share, number of holders, number of active addresses (past 30 days), monthly transfer count, monthly transfer volume (USD), 7-day and 30-day APY if listed, management fee, issuer/platform info, investor eligibility, tokenization type, and network distribution of holders/active addresses/volume.

Table \ref{tab:raw_variables} reports the key raw variables.  It shows, for example, that \texttt{BUIDL} has \$2.487B asset value with 108 holders and 30-days 24 active addresses, but monthly volume of only \$1.092B (8e-4 turnover).  By contrast, \texttt{BENJI} has a smaller \$823M asset value but 1106 holders and 17 active addresses, with only \$10M monthly volume. In the end, we exclude USDC from the main analysis and use it as a benchmark, since it is a large stablecoin with a market value of \$72 billion and approximately 42 million holders.

\begin{table}[htbp]
\centering
\caption{Raw RWA.xyz variables used in pilot metric construction}
\label{tab:raw_variables}
\scriptsize
\setlength{\tabcolsep}{3pt}
\begin{tabularx}{\textwidth}{lX
S[table-format=11.0]
S[table-format=8.0]
S[table-format=8.0]
S[table-format=13.0]
S[table-format=9.0]}
\toprule
Ticker & Category & {Asset value} & {Holders} & {Active addr.} & {Transfer vol.} & {Transfer count} \\
\midrule
BUIDL & Treasury & 2487654577 & 108 & 24 & 1092239594 & 88 \\
BENJI & Treasury & 823165231  & 1106 & 17 & 10063830   & 19 \\
OUSG  & Treasury & 612494992  & 55  & 15 & 110807065  & 34 \\
USTB  & Treasury & 721054020  & 99  & 21 & 285750617  & 709 \\
USDC  & Stablecoins & 72102491659 & 42268158 & 18527278 & 4204621116056 & 721150802 \\
USDY  & Treasury & 2143889939 & 14495 & 4325 & 566702395 & 158269 \\
HLSCOPE & Private credit & 4359598 & 45  & 6  & 123534    & 3 \\
STAC  & Structured credit & 101323883 & 4 & 1 & 3549447   & 1 \\
PAXG  & Gold & 4215207035 & 83949 & 11385 & 3765680650 & 263339 \\
XAUT  & Gold & 2591487591 & 56487 & 17589 & 4239912225 & 173584 \\
\bottomrule
\end{tabularx}
\end{table}

\begin{table}[htbp]
\centering
\caption{Variable definitions and orientation}
\label{tab:variables}
\scriptsize
\setlength{\tabcolsep}{3pt}
\renewcommand{\arraystretch}{1.20}
\begin{tabularx}{\textwidth}{lXXX}
\toprule
Variable & Definition & Direction of risk & Motivation \\
\midrule
Turnover 
& $Turnover_i =  \frac{TransferVolume_{30,i}}{AssetValue_i}$ 
& Lower turnover $\Rightarrow$ higher risk 
& Exitability proxy \\

Active ratio 
& $AR_i = \frac{ActiveAddresses_{30,i}}{Holders_i}$ 
& Lower ratio $\Rightarrow$ higher risk 
& Breadth of actual use \\

Transfer intensity 
& $TI_i = \frac{TransferCount_{30,i}}{Holders_i}$ 
& Lower intensity $\Rightarrow$ higher risk 
& Persistence of activity \\

Average transfer size 
& $ATS_i = \frac{TransferVolume_{30,i}}{TransferCount_{30,i}}$ 
& Larger size $\Rightarrow$ higher risk 
& Blocky rather than broad activity \\

Average value per holder 
& $AVH_i = \frac{AssetValue_i}{Holders_i}$ 
& Larger value $\Rightarrow$ higher risk 
& Concentrated ownership proxy \\

Holder HHI 
& $HHI_i = \sum_h s_{ih}^2$ 
& Higher HHI $\Rightarrow$ higher risk 
& Ownership concentration \\

Network concentration 
& $NHHI_i = \sum_c \omega_{ic}^2$ 
& Higher concentration or thin fragmentation can increase risk 
& Cross-chain distribution quality \\
\bottomrule
\end{tabularx}
\end{table}

\section{Three-Dimensional Explainable RWA Scoring Framework}
We construct three risk scores $L_i,C_i,M_i$ (higher = more risk) as follows:

\subsection{Liquidity Risk ($L$)}
$L$ combines turnover, active ratio $AR$, transfer intensity $TI$, and average transfer size $ATS$.  Intuitively, low turnover or low active ratio implies higher liquidity risk.  Formally, we first normalize each metric to a 0--100 risk scale by min-max scaling across the 10 assets:
\[
\text{RiskNorm}(x_i) = 
\begin{cases}
100\frac{x_i - \min(x)}{\max(x)-\min(x)}, & \text{if $x$ is risk-increasing (e.g.\ large ATS means risk)},\\
100\frac{\max(x) - x_i}{\max(x)-\min(x)}, & \text{if $x$ is protective (e.g.\ higher turnover lowers risk)}.
\end{cases}
\]
Concretely, turnover, active ratio, and transfer intensity are treated as \emph{protective} (higher is lower risk), while average transfer size is risk-increasing (higher block trades = higher risk).  We calculate their risk-normalized scores and then average them:
\[
\begin{aligned}
L_i =
\frac{1}{4}\Big(
&\mathrm{RiskNorm}(\text{Turnover}_i)
+ \mathrm{RiskNorm}(\text{ActiveRatio}_i)
\\
&+ \mathrm{RiskNorm}(\text{TransferIntensity}_i)
+ \mathrm{RiskNorm}(\text{ATS}_i)
\Big).
\end{aligned}
\]
For example, \texttt{BENJI} has very low turnover (0.012) and active ratio (0.015), yielding high $L$ (low liquidity).  In contrast, \texttt{XAUT} has the highest turnover (1.636), moderate active ratio (0.311), so its $L$ is moderate. 

\subsection{Concentration Risk ($C$)}
$C$ measures holder concentration.  We use number of holders (protective: more holders = lower risk), average value per holder (AVH = AssetValue/Holders, risk-increasing if high), and holder network concentration (NHHI) as proxies.  Again we min-max normalize:
\[
C_i = \frac{ \mathrm{RiskNorm}(\text{Holders}_i) + \mathrm{RiskNorm}(\text{AVH}_i) + \mathrm{RiskNorm}(\mathrm{NHHI}_i) }{3}.
\]
Here $\mathrm{RiskNorm}(\text{Holders})$ uses the protective formula (fewer holders = higher risk).  Since network HHI is already a concentration (0 to 1), we scale it to 0--100 (risk increases with HHI).  For example, \texttt{BENJI} has 1106 holders (many compared to \texttt{BUIDL}) so one component is low risk, but its network HHI is 1 (all on Stellar) giving high risk on that dimension.  We average all three contributions. 

\subsection{Market-Quality Risk ($M$)}
$M$ captures on-chain trading concentration and fragmentation.  We use the Herfindahl index of active addresses by chain and of transfer volume by chain.  Both of these are risk-increasing (more concentrated chains $\to$ higher $M$).  Define 
\[
M_i = \frac{\mathrm{RiskNorm}(\mathrm{HHI}_{\mathrm{active},i}) + \mathrm{RiskNorm}(\mathrm{HHI}_{\mathrm{volume},i})}{2}.
\]
E.g.\ \texttt{HLSCOPE} trades mostly on Polygon (active HHI=1, volume HHI=1) so $M=100$.  By contrast, \texttt{USDC} is widely distributed (very low HHIs), so $M\approx0$.  Chain fragmentation (many small-chain trades) effectively lowers $M$. 

The summary of the variables used in this paper can be seen in Table~\ref{tab:variables}.

\subsection{Composite Score}
We combine the three sub-scores into a single composite risk index.  The baseline composite is the simple average:
\[
\text{Composite}_i = \frac{L_i + C_i + M_i}{3}.
\]
We also analyze sensitivity by alternative weights: (i) {\it liquidity-heavy} $(0.5,0.25,0.25)$, (ii) {\it concentration-heavy} $(0.25,0.5,0.25)$, and (iii) {\it market-quality-heavy} $(0.25,0.25,0.5)$.  Higher composite means empirically higher risk (low liquidity, high concentration or poor market quality).


\section{Empirical Results}

Table~\ref{tab:calculated_metrics_corrected} reports the derived variables used to construct the empirical liquidity, concentration, and market-quality scores. The results show substantial heterogeneity across tokenized RWA products. USDC records the highest turnover and transfer intensity, reflecting its role as a highly active stablecoin benchmark rather than a fund-type tokenized asset. Among the RWA products, XAUT and PAXG show relatively high turnover, suggesting stronger observed transfer circulation than most tokenized Treasury and credit products. By contrast, BENJI, HLSCOPE, and STAC exhibit low turnover and low transfer intensity, indicating limited observed secondary activity relative to their asset value and holder base. The average value per holder also varies sharply, with BUIDL, OUSG, USTB, and STAC showing much higher values than commodity tokens, which suggests stronger institutional concentration. Finally, the NHHI values indicate that several products remain highly concentrated across networks, with BENJI, HLSCOPE, STAC, and PAXG recording an NHHI of 1.0000, while USDC and USDY show more dispersed network activity. Overall, the table supports the papers main argument that asset value alone is insufficient to assess tokenized RWA quality because liquidity, holder breadth, and market activity differ substantially across products.

\begin{table}[htbp]
\centering
\caption{Corrected calculated liquidity, concentration, and market quality variables}
\label{tab:calculated_metrics_corrected}
\scriptsize
\setlength{\tabcolsep}{3pt}
\begin{tabular}{llrrrrr}
\toprule
Ticker & Category & Turnover & Active ratio & Transfer intensity & AVH & NHHI \\
\midrule
BUIDL   & Treasury          & 0.4391  & 0.2222 & 0.8148  & 23033839 & 0.4939 \\
BENJI   & Treasury          & 0.0122  & 0.0154 & 0.0172  & 744273   & 1.0000 \\
OUSG    & Treasury          & 0.1809  & 0.2727 & 0.6182  & 11136273 & 0.6626 \\
USTB    & Treasury          & 0.3963  & 0.2121 & 7.1616  & 7283374  & 0.8141 \\
USDC    & Stablecoins       & 58.3145 & 0.4383 & 17.0613 & 1706     & 0.1414 \\
USDY    & Treasury          & 0.2643  & 0.2984 & 10.9189 & 147905   & 0.2660 \\
HLSCOPE & Private credit    & 0.0283  & 0.1333 & 0.0667  & 96880    & 1.0000 \\
STAC    & Structured credit & 0.0350  & 0.2500 & 0.2500  & 25330971 & 1.0000 \\
PAXG    & Gold              & 0.8934  & 0.1356 & 3.1369  & 50212    & 1.0000 \\
XAUT    & Gold              & 1.6361  & 0.3114 & 3.0730  & 45878    & 0.9050 \\
\bottomrule
\end{tabular}
\end{table}

Table \ref{tab:scores} shows the computed $L,C,M$ and composite scores for the 10 assets.  Large institutional issuances like \texttt{PAXG} (gold) have very high holder counts and multi-chain activity, yielding relatively low $L$ and moderate $C$ risk.  For instance, \texttt{PAXG} achieves $L\approx62.96$ (lower risk) due to very high turnover (many transfers) and decent active participation, even though it has only one chain of issuance (raising some concentration).  Conversely, \texttt{BENJI} (short-duration bond token) has low turnover (0.012) and tiny active ratio (1.5\%), giving it a high liquidity risk $L\approx76.06$, despite having many holders which slightly moderates $C$. 

Although \texttt{USDC} is included in the raw variable table as a benchmark, it is excluded from the final RWA scoring sample because it is a reserve-backed stablecoin rather than a fund-type tokenized real-world asset. This distinction is important because \texttt{USDC} has a substantially different market function, user base, and transfer profile from tokenized Treasury, credit, and commodity products. If the same liquidity, concentration, and market-quality variables are applied mechanically, \texttt{USDC} receives near-zero risk scores because it has very high turnover, a large active-address base, low average value per holder, and broad cross-chain usage. This result is useful as a benchmark because it shows what a highly active and widely distributed on-chain asset looks like under the proposed metrics. However, including \texttt{USDC} in the main RWA sample would distort the normalization because its transfer activity is orders of magnitude larger than the other products. For this reason, \texttt{USDC} is reported as a benchmark but excluded from the final composite RWA ranking.

In contrast, private-credit tokens \texttt{STAC} and \texttt{HLSCOPE} score highest.  \texttt{STAC} has only 4 holders and minimal trading (only one transfer recently), giving $L\approx67.92$, $C=100$, $M=100$, and composite $\approx89.3$.  Even adjusting weights, \texttt{STAC} remains at the top (see Table \ref{tab:sensitivity}).  \texttt{HLSCOPE} similarly has very few active trades, so both $L$ and $C$ are high.  

\begin{table}[htbp]
\centering
\caption{Empirical pilot scores for tokenized RWAs using RWA.xyz data. Higher values indicate higher observed risk under the liquidity, concentration, and market-quality framework. The table is sorted by the equal-weight composite score in descending order.}
\label{tab:scores}
\begin{tabular}{p{2.0cm}p{2.8cm}rrrr}
\toprule
Ticker & Category & $L$ & $C$ & $M$ & Composite (Equal) \\
\midrule
STAC    & Structured credit          & 67.92 & 100.00 & 100.00 & 89.31 \\
BENJI   & Treasury                   & 76.06 & 67.64  & 100.00 & 81.23 \\
HLSCOPE & Private credit             & 68.02 & 66.79  & 100.00 & 78.27 \\
PAXG    & Gold                       & 62.96 & 66.66  & 100.00 & 76.54 \\
USTB    & Treasury                   & 64.58 & 81.41  & 83.01  & 76.50 \\
XAUT    & Gold                       & 52.37 & 62.99  & 92.39  & 69.26 \\
OUSG    & Treasury                   & 65.39 & 68.22  & 57.68  & 63.77 \\
BUIDL   & Treasury                   & 86.42 & 77.33  & 22.04  & 61.93 \\
USDY    & Yield-bearing Treasury     & 95.90 & 40.87  & 42.54  & 59.77 \\
\bottomrule
\end{tabular}

\vspace{0.2cm}
\begin{minipage}{0.95\textwidth}
\footnotesize
\textit{Notes:} $L$ denotes liquidity risk, $C$ denotes concentration risk, and $M$ denotes market-quality risk. 
All scores are normalized on a 0 to 100 scale, where higher values indicate higher observed risk. 
The composite score is calculated using equal weights across the three dimensions. 
USDC is excluded from the final RWA scoring sample because it is used as a stablecoin benchmark rather than as a fund-type tokenized real-world asset.
\end{minipage}
\end{table}

\begin{table}[htbp]
\centering
\caption{Sensitivity of composite scores to alternative weighting schemes. Higher values indicate higher observed risk, and the table is sorted by the equal-weight composite score in descending order.}
\label{tab:sensitivity}
\begin{tabular}{p{2.0cm}rrrr}
\toprule
Ticker & Equal weights & Liquidity-heavy & Concentration-heavy & Market-quality-heavy \\
\midrule
STAC    & 89.31 & 83.96 & 91.98 & 91.98 \\
BENJI   & 81.23 & 79.94 & 77.84 & 85.93 \\
HLSCOPE & 78.27 & 75.71 & 75.42 & 83.70 \\
PAXG    & 76.54 & 73.15 & 74.07 & 82.41 \\
USTB    & 76.50 & 72.29 & 78.26 & 79.85 \\
XAUT    & 69.26 & 65.03 & 67.69 & 75.03 \\
OUSG    & 63.77 & 64.17 & 64.88 & 62.24 \\
BUIDL   & 61.93 & 68.05 & 65.78 & 51.96 \\
USDY    & 59.77 & 60.07 & 59.67 & 60.01 \\
\bottomrule
\end{tabular}

\vspace{0.2cm}
\begin{minipage}{0.95\textwidth}
\footnotesize
\textit{Notes:} The equal-weight composite gives one-third weight to liquidity risk, concentration risk, and market-quality risk. 
The liquidity-heavy specification uses weights of 0.50, 0.25, and 0.25 for $L$, $C$, and $M$, respectively. 
The concentration-heavy specification uses weights of 0.25, 0.50, and 0.25. 
The market-quality-heavy specification uses weights of 0.25, 0.25, and 0.50. 
USDC is excluded from the final RWA scoring sample because it is used only as a stablecoin benchmark.
\end{minipage}
\end{table}

\section{Discussion}
Our results underscore the pitfalls of using TVL alone to assess RWA quality.  For instance, \texttt{USTB} and \texttt{BENJI} are both U.S. Treasury funds of similar asset value (\$0.7–0.8B) but have very different profiles.  \texttt{BENJI} has many holders but very little transfer activity, leading to high $L$ and moderate $C$ risk.  \texttt{USTB} trades more frequently (turnover 0.396 vs 0.012) and thus has lower $L$, but fewer holders and a concentrated investor base, giving higher $C$.  In our pilot, no asset has high scores on all dimensions except \texttt{STAC}, reflecting its very low liquidity and high concentration.

Gold-backed tokens (\texttt{PAXG}, \texttt{XAUT}) show interesting patterns: they have large asset values and many holders, but relatively low monthly volume compared to assets like treasury funds.  This yields moderate $L$ (gold markets are somewhat active) but high $M$ due to concentration of volume on single chains.  Private credit funds (\texttt{HLSCOPE}, \texttt{STAC}) have small TVL and extremely low trading, so their liquidity risk is very high.

Qualitatively, higher $L$ indicates assets where exiting could be hard.  Higher $C$ indicates potential vulnerability to large-holder actions.  Higher $M$ indicates reliance on narrow trading venues or chains.  

\section{Implications}
For \emph{investors}, our findings imply that beyond shiny metrics like market cap, one should scrutinize actual trading data.  A high-TVL RWA might still trap capital if no one trades it.  Our framework offers a quick quantitative check using publicly available data.  For \emph{issuers}, the message is to build genuine market support: encourage broad participation and continuous trading (e.g.\ via AMMs or approved exchanges), not just focus on asset raise.  For example, a token with poor $L$ could lower fees or provide liquidity incentives.  \emph{Regulators} should note that transparency tools (like RWA.xyz) can reveal secondary-market health, but they do not replace due diligence on legal redemption or custody promises.  Regulators might mandate disclosure of turnover and holder metrics for token offerings.  Finally, \emph{data providers} should standardize the key metrics we used (transfer counts, holder counts, active addresses, chain splits) so that all market participants can apply similar scoring.

\section{Limitations and Future Research}
This pilot uses only public on-chain data.  We do \emph{not} capture off-chain trades or locked-of records.  Some assets may have hidden liquidity (e.g.\ private bilateral trades) not seen on-chain.  Metrics like holder count ignore beneficial owners and do not capture whether many sub-accounts belong to the same entity.  Our network HHIs use chain-level aggregates; wallet-level concentration requires proprietary or onchain analysis beyond RWA.xyz.  Also, factors like legal enforcement, redemption terms, custody risk, or smart-contract safety are omitted here.  These are important RWA risks (e.g.\ stablecoin issuers flag reserve risk), but they are not quantifiable from the available data.  Future work should combine on-chain liquidity metrics with off-chain disclosures (issuer reports, redemption histories) and potentially supervised learning when enough risk events (defaults, liquidity crises) have occurred.  Extending this scoring to a time series (monthly or quarterly panel) would also allow testing how changes in $L,C,M$ predict actual market outcomes.

\section{Conclusion}
This paper develops an explainable, three-dimensional scoring framework for evaluating tokenized real-world assets using publicly observable RWA.xyz market, holder, and activity data. Rather than treating asset value or TVL as sufficient indicators of quality, the framework focuses on three empirically measurable dimensions: liquidity risk, concentration risk, and market-quality risk. The pilot results show that tokenized RWA products differ substantially in observed tradability, holder breadth, average value per holder, and network concentration. In particular, some products with meaningful asset value still show limited transfer intensity, concentrated holder structures, or weak market activity, suggesting that tokenization alone does not automatically create exitability.

The findings have practical implications for investors, issuers, data providers, and regulators. Investors should evaluate whether a tokenized asset can actually be traded or exited, not only whether it has large reported asset value. Issuers should disclose activity, holder, transfer, and network-distribution metrics alongside AUM or NAV. Regulators and market-data providers may also benefit from standardized public reporting of turnover, active participation, transfer intensity, and ownership concentration proxies. The framework remains a public-data pilot and does not capture legal enforceability, redemption rights, custody arrangements, reserve verification, or smart-contract governance. Future research should extend the model using panel data, wallet-level holder distributions, redemption records, exchange liquidity, and issuer-level documentation to provide a more complete assessment of tokenized RWA risk.

\end{document}